\documentclass[preprintnumbers,amsmath,amssymb,floatfix,10pt,prd,onecolumn,layout,
superscriptaddress,nofootinbib]{revtex4}
\usepackage{latexsym}
\usepackage{epsfig}
\usepackage{epstopdf}
\usepackage{graphicx}
\usepackage{amssymb}
\usepackage{amsmath}
\usepackage{dcolumn}
\usepackage{bm}
\usepackage{color}
\usepackage{comment}

\begin{document}

\title{\bf Constant-roll Warm Inflation within Rastall Gravity}
\author{Rabia Saleem}
\altaffiliation{rabiasaleem@cuilahore.edu.pk}
\author{Iqra Shahid}
\altaffiliation{ikra.2016@gmail.com}
\author{M. Israr Aslam}
\altaffiliation{mrisraraslam@gmail.com}
\author{Abdul Wahab}
\altaffiliation{abwahab927@gmail.com}
\affiliation{Department of Mathematics, COMSATS University Islamabad, \\Lahore Campus, Pakistan.}

\begin{abstract}
This research paper used a newly proposed strategy for finding the exact inflationary solutions to the Friedman equations in the context of Rastall theory of gravity (RTG), which is known as constant-roll warm inflation (CRWI). The dissipative effects produced during WI are studied by introducing a dissipation factor $Q=\frac{\Gamma}{3H}$, where $\Gamma$ is the coefficient of dissipation. We establish the model to evaluate the inflaton field, effective potential requires to produce inflation, and entropy density. These physical quantities lead to developing the important inflationary observables like scalar/tensor power spectrum, scalar spectral index, tensor-to-scalar ratio, and running of spectral-index for two choices of obtained potential that are $V_0=0$ and $V_0\neq0$. In this study, we focus on the effects of the theory parameter $\lambda$, CR parameter $\beta$, and dissipation factor $Q$ (under a high dissipative regime for which $Q=$constant) on inflation, and are constrained to observe the compatibility of our model with Planck TT+lowP (2013), Planck TT, TE, EE+lowP (2015), Planck 2018 and BICEP/Keck 2021 bounds. The results are feasible and interesting up to the $2\sigma$ confidence level. Finally, we conclude that the CR technique produces significant changes in the early universe.\\

{\bf Keywords:} Modified Gravity; Warm Inflation; Constant-roll Approximation.
\end{abstract}
\date{}
\maketitle
\section{Introduction:}

One of the proposed phenomena for understanding the early evolution of the universe is inflation.
It states that the universe expands at an extraordinarily fast rate in a very short period of time.
Starobinsky \cite{1} was the first one to propose such an expansion theory in early times,
and Guth \cite{2} was the first to use it as a possible solution to the problems of the big-bang theory.
The scenario has been investigated extensively and modified in many different ways \cite{3}-\cite{12}
due to its high compatibility with cosmological data \cite{13}. Generally, in many inflationary models,
inflation is caused by a scalar field, known as inflaton. To ensure enough inflation, the potential of
the inflaton field is usually chosen to be flat so that the scalar field slowly descends from its maximum
potential to its minimum, i.e., the slow-roll (SR) condition is satisfied. The SR parameters $(\epsilon, \eta)$,
which should be very small during an inflationary period, perfectly describe this behavior. A relatively flat
potential provides the smallness of these parameters \cite{14}-\cite{17}.

Inflationary cosmology explains the chaotic inflation \cite{18}, hybrid inflation \cite{19} and new inflation
\cite{20} that are based on the supposition that the scalar field $\phi$ driving inflation satisfies certain
SR conditions. In general relativity (GR), the Klein-Gordon equation of motion under SR approximation is as follows
\begin{equation}\label{1.1}
\ddot{\phi}+3H\dot{\phi}+\frac{\partial V}{\partial \phi}=0,
\end{equation}
where $H$ represents the Hubble expansion and $V(\phi)$ is the effective potential of inflaton. In comparison
with the other terms, the second derivative of $\phi$ is negligible and left with a first-order differential
equation in $\phi$, $3H\dot{\phi}+\frac{\partial V}{\partial \phi}=0$. For the system, $M_p^2=1$, the following
assumptions describing the flatness of the potential are generally similar to the slow evolution of $\phi$.
\begin{equation}\label{1.2}
\epsilon=-\frac{\dot{H}}{H^{2}}\simeq\frac{1}{2}\bigg(\frac{\frac{\partial V}{\partial \phi}}{V}\bigg)^{2}\ll 1,\qquad
\eta=\bigg|\frac{\ddot{\phi}}{H\dot{\phi}}\bigg|\simeq \bigg|\frac{\frac{\partial^2 V}{\partial \phi^2}}{V}\bigg|\ll 1.
\end{equation}
Outside the SR regime, one can get a quasi-de-Sitter rate of expansion by assuming the SR solution of the field equations
and then employing a scalar field $\phi$ with an approximately flat potential, as described in \cite{21}. The condition
$\ddot{\phi}+3H\beta\dot{\phi}=0$, can be used to describe the cases of SR and beyond SR evolution. The values $\beta=0$
and $\beta=1$ represent the cases of ultra SR (USR) and beyond SR, respectively. In the USR inflation, the SR condition
$(\eta\ll1)$ is violated.

If the potential of the scalar field is very flat, then the acceleration of the inflaton is locked by the friction term
and the slow-roll parameter $\eta\sim3$, this is the USR inflation \cite{47a,47b}. Recently, the USR inflation was
generalized to CR inflation with the SR parameter $\eta$ being a constant \cite{21,22}. The CR inflation has richer physics
than the SR inflation does. For example, it can generate large local non-Gaussianity and the curvature perturbation may
grow on the super-horizon scales \cite{21,22,22a}. The concept of CR inflation is to assume a continuous spectrum of
$\beta$ \cite{22}-\cite{26}, which can be implemented in modified theories of gravity \cite{27}-\cite{29}. The CR
inflation has recently received a lot of attention so one can observe a large number of inflationary models in this approach
\cite{31}-\cite{40}. Furthermore, the idea of USR inflation was used to generate the primordial black holes \cite{R1,R2},
and a short period of USR inflation with small $\epsilon$ can be the mechanism for producing the primordial black holes in
a similar way to the chaotic new inflation model \cite{R3}.

Warm inflation was proposed in 1995 \cite{41aa} to resolve the issue of reheating the inflaton.
In this scenario, the scalar field is still thought to be the primary source of inflation and
it still fluctuates slowly, but there are notable differences between WI and cold inflation (CI).
The first and foremost difference is the production of radiation during WI, so it can interact
with other fields during the inflationary period. Due to interaction, energy transmits from
inflaton field to radiation field and at the end of inflation, the universe remains hot. The
second difference is that the reheating mechanism is not necessary. Another difference is related
to fluctuations. Contrary to CI (where the quantum fluctuations dominates over thermal fluctuations
\cite{16,41b}), and thermal fluctuations dominates if the condition $T>H$ ($T$ is the temperature)
is fulfilled \cite{41}-\cite{44}.

In the evolution of the universe, dissipative effects including shear and bulk viscosity are of considerable significance.
A dissipation term in the energy conservation equation of inflaton mediates the transfer of inflaton's energy to the
radiation bath during WI. Because of inner couplings in the radiation fluid, an additional effect is also observed.
The internal dissipation in the radiation fluid can slightly disturb the thermal equilibrium. The radiation fluid behaves
as a non-ideal fluid and as a result, the viscosity effects get significant. The only viscous effect that appears in the
background equations is the bulk viscous pressure usually denoted by $\Pi$ \cite{bulk}. Due to the decay of massive particles
within the fluid, this process is an entropy-producing scalar phenomenon. On the other hand, $\Pi$ has
entropy-producing property. In cosmology, especially in terms of inflation, the investigation of the $\Pi$ effect is
centered on the $\Pi$ as a negative pressure. Since it acts as a starting point of the acceleration in the cosmic
expansion, there has been huge interest by scholars in discussing the effects of $\Pi$. Detailed analysis of the
dynamics of WI with bulk viscous pressure is presented in \cite{45}. Motohashi et al. \cite{21} investigated an inflationary
scenario where the rate of roll defined by $\ddot{\phi}/H\dot{\phi}=-(3+\alpha)$, remained constant and showed that all exact
solutions are satisfying the constant rate of roll ansatz. Gao \cite{47} explored the observational constraint on CR inflation
and deduced that the results of $n_{s}-r$ are different from the SR inflation \cite{R4}. Another model presented by Gao where
the effects of Gauss-Bonnet inflation with a constant rate of roll is studied. Kamali et al. \cite{36} introduced a new concept
by applying CR approximation on the model with WI in high dissipative regime and found that $n_{s}-r$ trajectories remained
compatible with Planck 2018 data. Awad et al. \cite{awad} investigated the implications of the CR condition on the inflationary
era in $f(T)$ gravity. They also analyzed the phase space of the CR inflation in teleparallel gravity and discussed the physical
significance of the resulting fixed points and trajectories. It is concluded that the resulting theory can be compatible with the
current observational data for a wide range of parametric values.

General Relativity can be modified by introducing a minimal/non-minimal coupling between matter fields and geometry. Since in modified gravities, the energy-momentum tensor (EMT) does not conserve because of the divergence-free tensor field, which is coupled with geometry in a minimal way \cite{33n}-\cite{34n}. Modified theories of gravity include RTG, which is considered the classical formulation of particle production in cosmology \cite{35n}. The covariant divergence of EMT $T_{\nu;\mu}^{\mu}=0$, is not sound good anymore \cite{36n}-\cite{39n}. In 1972, this concept was first published by Peter Rastall \cite{39n} who proposed a modified gravitational model in which $T_{\nu;\mu}^{\mu} \propto R_{,\nu}$, where $R$ is the curvature invariant and the proportionality constant $\lambda$ is called  Rastall parameter, i.e., $T_{\nu;\mu}^{\mu}=\lambda R_{,\nu}$, this leads to the recovery of conservation law in flat space-time when $\lambda=0$. Capone et al. \cite{40n} studied the possibility that the observed cosmological expansion was possibly restored under RTG. Nojiri et al. \cite{28} worked on the CR inflation in the framework of $f(R)$ gravity. Some cosmological consequences are explored in the context of generalized RTG by Das et al. \cite{41n}. Saleem and Hassan \cite{43n} worked on the dynamics of WI encouraged by vector field in the context of RTG and deduced the compatibility of this modified theory with Planck observational data published in 2018. Saleem and Shahid \cite{me} investigated WI through irreversible thermodynamics of open systems along with matter creation/decay within RTG.

The study of the inflationary phenomenon under CR conditions seems to be the center of attraction these days. Authors in Ref. \cite{35} applied inflationary models using the concept of CR that was previously established in GR and $f(R)$ gravity. In the presence of multi-scalar fields, CR inflation is analyzed in \cite{multi}. The authors in \cite{s2} examined a quintessence scalar field for WI with a constant dissipative coefficient that led to the CR inflationary model in the framework of brane-world cosmology. Motivating by this CR technique \cite{36} of finding exact inflationary solutions as a function of e-foldings $(N)$ in the presence of radiation, we aim to apply the technique on RTG. So, we establish basic formalism for CR inflation in \textbf{section II}. Here we extend the work to discuss WI under the CR limit and find the exact solutions of the dynamical equations in terms of $N$ in RTG in \textbf{section III}. In \textbf{section IV}, we investigate the perturb parameters such as scalar/tensor power spectra $P_\xi$, scalar spectral index $n_{s}$,  its running $\alpha_{s}$ and tensor-to-scalar ratio $r$ in high dissipative regime and show that the results of $R,~n_{s},~\alpha_{s}$  are in good consistency with Planck observational data under constrained values of the involved model parameters $\lambda,~\beta$ and $Q$. We summarized our obtained results in \textbf{section V}.

\section{Basic Formalism of Constant-Roll Inflation}

Let us assume that an inflationary model with a homogeneous scalar field $\phi$ and the metric of the
early universe is a flat Friedman Robertson Walker (FRW) model. The field equations for RTG are written
as \cite{39n}
\begin{equation}
G_{\mu\nu}+\kappa\lambda g_{\mu\nu}R=\kappa T_{\mu\nu}.
\end{equation}
The corresponding field equations for flat FRW within RTG are given below
\begin{eqnarray}\label{2.1}
(12\kappa\lambda-3)H^2+6\kappa\lambda\dot{H}=\kappa\rho, \\\label{2.2}
(12\kappa\lambda-3)H^2+(6\kappa\lambda-2)\dot{H}=\kappa P,
\end{eqnarray}
where $\rho$ and $P$ are the total energy density and pressure of the fluid. For $\lambda=0$, the Einstein field equations
will be the same as obtained for flat FRW in GR. The Bianchi identity $G^{\mu\nu}_
{;\mu}=0$, leads to following equation of continuity \cite{43n}
\begin{equation}\label{2jj}
\bigg(\frac{3\kappa\lambda-1}{4\kappa\lambda-1}\bigg)\dot{\rho}+\bigg(\frac{3\kappa\lambda}{4\kappa\lambda-1}\bigg)
\dot{P}+3H(\rho+P)=0.
\end{equation}
The fluid is considered to be composed of scalar filed with energy density $\rho_\phi=\frac{1}{2}\dot{\phi}^2+V(\phi)$
and pressure $P_\phi=\frac{1}{2}\dot{\phi}^2-V(\phi)$ along with radiation
having energy density $\rho_r$ and pressure $P_r=(\gamma-1)\rho_r$ ~$(\frac{2}{3}\leq\gamma\leq2)$. The corresponding conservation equations become
\begin{eqnarray}\label{M1}
\left(\frac{1-3\kappa\lambda\gamma}{1-4\kappa\lambda}\right)\dot{\rho_{\phi}}&+&3H(\rho_{\phi}+P_{\phi})=
0,\\\label{M2}
\left(\frac{1-3\kappa\lambda}{1-4\kappa\lambda}\right)\dot{\rho_{r}}&+&4H\rho_{r}=0,
\end{eqnarray}

In the case of flat potential $V'=0$, and according to equation of motion of inflaton field in GR, one has
$\ddot{\phi}=-3H\beta\dot{\phi}$. Then, the second SR parameter $\eta=3$ \cite{47a}, and it will no longer
small. This was the starting point of the scenario of CR inflation, where the smallness of the second SR
parameter was released. The CR condition using Eq.(\ref{M1}) can be modified in RTG as
\begin{equation}\label{2.3}
\ddot{\phi}=-3\beta C_1 H \dot{\phi},
\end{equation}
where $C_{1}=\frac{1-3\kappa\lambda\gamma}{1-4\kappa\lambda}$. For $\lambda=0$, it will reduce to standard CR condition.
On integrating the above equation with respect to $t$, we get equation of motion for inflaton as
\begin{equation}\label{2.4}
\dot{\phi}=\dot{\phi_{0}}a^{-\frac{3\beta}{C_1}},
\end{equation}
where $\dot{\phi_{0}}$ is the initial value of inflaton at $t_{i}$ and we consider it as $a(t_{i})=1$. In term of number
of e-folds, it can be defined as
\begin{equation}\label{2.5}
N=\log\frac{a}{a_{i}}=\log(a) \quad \Rightarrow a=e^{N}.
\end{equation}
Putting the above expression in Eq. (\ref{2.4}), we can modify that equation in terms of e-foldings
\begin{equation}\label{phi}
\dot{\phi}=\dot{\phi_{0}}e^{-\frac{3\beta}{C_1}N}.
\end{equation}
These equations are valid in the case of CR inflation. We can also modify these dynamical equations to discuss WI scenario
where a dissipation factor $Q=\frac{\Gamma}{3H}$, will be involved in the conservation equations.

\section{Warm Inflation Inspired by Constant-roll Condition}

Following the criteria given in \cite{36}, we will find the inflationary solutions of the modified field equation
under CR condition. The continuity Eqs. (\ref{M1}) and (\ref{M2}) with dissipation factor become
\begin{eqnarray}\label{M11}
C_{1}\dot{\rho_{\phi}}+3H(\rho_{\phi}+P_{\phi})&=&
\Gamma(\rho_{\phi}+P_{\phi}),\\\label{M22}
C_{1}\dot{\rho_{r}}+4H\rho_{r}&=&\Gamma(\rho_{\phi}+P_{\phi}).
\end{eqnarray}
The above equations can be further modified as using standard $\rho_\phi$ and $P_{\phi}$
\begin{eqnarray}\label{3.1}
C_{1}\ddot{\phi}+C_{1}V'(\phi)+3H(1+Q)\dot{\phi}&=&0,\\\label{3.2}
C_{1}\dot{\rho_{r}}+4H\rho_{r}&=&\Gamma \dot{\phi}^{2}.
\end{eqnarray}
In WI, we want to achieve the non-zero temperature of the cosmos. Therefore, we consider that inflaton can dissipate into relativistic degrees of freedom. Warm inflation with high dissipation ($\Gamma\gg3H$) allows the quasi-de-Sitter universe and inflaton to evolve slowly even though the potential does not fulfill the SR conditions. In this regard, many theories have a chance to be compatible with the observational data because this impact relaxes the restrictions on the potential required to achieve inflation. Swampland theory is another inspiration of WI, which shows that inflation can be compatible with swampland theory when $Q\gg1$ (or $Q=$constant) \cite{49}.

Since the CR expansion is required, therefore Eqs. (\ref{2.3}) and (\ref{2.4}) remain valid. Equation (\ref{2.3}) can still be characterized with SR approximation as for $\beta\ll1$, that is identical to SR case. From the above system of equations, our aim is to extract the evolution of the inflaton field and the expression of required potential. Now using Eqs. (\ref{2.4}) and (\ref{3.2}), we get radiation density as
\begin{equation}\label{3.3}
\rho_{r}=TS=\frac{3Q\dot{\phi_{0}}^{2}}{2(2-3C_{1}\beta)}a^{-\frac{3\beta}{C_1}}+\rho_{r_{0}}a^{-\frac{4}{C_{1}}},
\end{equation}
where $\rho_{r_{0}}$ is the current value of radiation energy density and $S$ be the entropy density. Above solution can be characterized into two cases:
\begin{itemize}
\item When $\beta\ll1$, it shows consistency with standard SR WI. In this case, one can find approximately constant value of $\rho_{r}$.

\item When $\rho_{r}$ deviates from the SR condition, it is suppressed strongly and exponentially. Thus inflationary phase occur, containing elements of both CI and WI. The cosmic friction is increased (a special characteristic of WI), while the temperature is extremely close to absolute zero (as in CI). The right-hand side of Eq. (\ref{3.3}) is proportional to $Q$ and becomes negative for $\beta>\frac{2}{3C_{1}}$.
As a result, even if the inflaton's energy density dissipates completely into radiation density, one could mistakenly assume that for $\beta>\frac{2}{3C_{1}}$, the radiation receives smaller energy density (and thus lower temperature) in the case of CI. In WI scenario, temperature decreases slowly as compared to the CI scenario. For this, we assume an unsourced radiation term $\tilde{\rho_{r}}=\tilde{\rho_{r_{_{0}}}}a^{-\frac{4}{C_1}}$, and $\rho_{r}$ term from Eq. (\ref{3.3}). So, the ratio of
$\frac{\rho_{r}}{\tilde{\rho_{r}}}$ always increasing as
\begin{equation}\label{3.4}
\frac{d}{dt}\bigg(\frac{\rho_{r}}{{\tilde\rho_{r_{0}}a^{-\frac{4}{C_{1}}}}}\bigg)=3\frac{\dot{\phi_{0}}^{2}}{\tilde{\rho_{r_{0}}}}C_{1}HQ a^{\frac{1}{C_{1}} (4- 6\beta)}>0, \quad \forall \beta.
\end{equation}
Since there exists a term in radiation, which possibly be negative, as if one assume $N$
small enough that  leads to negative radiation energy density, which is nonphysical.
To this regard, we restrict ourselves to the case
\begin{equation}\label{N}
N>\frac{C_{1}}{2(2-\frac{3}{2}\beta)}\log\bigg(\frac{2\rho_{r_{0}}(3C_{1}\beta-2)}{3Q \dot{\phi_{0}}^{2}}\bigg),
\end{equation}
which provides $\rho_{r}>0$. Once more, this restriction on $N$ is compulsory for $\beta>\frac{2}{3 C_{1}}$,
as for $\beta<\frac{2}{3 C_{1}}$ one can find $\rho_{r}>0$, for all $N$.
\end{itemize}

To solve Eqs. (\ref{2.1}) and (\ref{2.2}), we used Eq. (\ref{3.3}), which leads to
\begin{equation}\label{3.5}
-M_{p}^{2}\frac{d}{dN}(H^{2})=\dot{\phi_{0}}^{2} \left(\frac{2Q+2-3C_{1}\beta}{2-3C_{1}\beta}\right) e^{-\frac{6\beta}{C_1} N}+\frac{4}{3} \rho_{r_{0}} e^{-\frac{4N}{C_{1}}}.
\end{equation}
Now by applying integration on the both sides of Eq. (\ref{3.5}) with respect to
$N$, we get the following equation
\begin{equation}\label{3.6}
3M_{p}^{2}H^{2}=\frac{\dot{\phi_{0}}^{2}}{2\beta} \left(\frac{2Q+2-3C_{1}\beta}{2-3C_{1}\beta}\right) e^{-\frac{6\beta}{C_1} N}+C_{1} \rho_{r_{0}} e^{-\frac{4N}{C_{1}}}+V_{0}.
\end{equation}
Now for $Q>0$, this expression remains positive for $\beta<\frac{2}{3C_{1}}$ or for $\beta>\frac{2(1+Q)}{3C_{1}}$. Let us assume $\rho_{r_{0}}=V_{0}=0$ to obtain $\epsilon=3\beta$ (in GR for $\lambda=0$), which set $\beta\ll1$. It is also compatible with CI scenario with equation $\epsilon=3\beta$. We can evaluate the potential $V$ in terms of e-folds with the help of energy density by combining Eqs. (\ref{2.1})
and (\ref{3.6}) as under
\begin{equation}\label{3.7}
V=V_{0}+\frac{\dot{\phi_{0}}^{2}}{2\beta}\bigg[1+C_{2}Q-\beta \bigg] e^{-\frac{6\beta}{C_1} N},
\end{equation}
where $C_{2}=\frac{2-3\beta}{2-3C_{1}\beta}$. One can obtain the range of $\beta$ for the case of WI, i.e., $\beta\leq1+C_{2}Q$. Another
method of finding $V$ is through the effective equation of motion of $\phi$, which is in the following form
\begin{equation}\label{3.8}
V'(\phi)+3HC_{1}\dot{\phi}\bigg(\frac{1+Q}{C^2_{1}}-\beta\bigg)=0.
\end{equation}
It is worth mentioning that for $\beta=\frac{1+Q}{C^2_{1}}$, one can find $V=V_0$. This scenario is completely different from CI under CR constraints. During high dissipation $Q\gg1$, one may get $\beta\gg1$, which increased the deviation from SR regime.

In order to obtain potential in terms of $\phi$, first we need to obtain $N=N(\phi)$. For this, we find the expression of $\phi(N)$ from Eq. (\ref{phi}) as
\begin{equation}\label{3.9}
\phi(N)=\pm\sqrt{3}\dot{\phi_{0}}M_{p}\int\frac{e^{-\frac{6\beta}{C_1} N}}{\sqrt{\frac{\dot{\phi_{0}}^{2}}{2\beta}\bigg(\frac{2Q+2-3C_{1}\beta}{2-3C_{1}\beta}\bigg)e^{-\frac{6\beta}{C_1} N}+C_{1}\rho_{r_{0}}e^{-\frac{4N}{C_{1}}}+V_{0}}} dN.
\end{equation}
It is not possible to find the solution of $\phi(N)$ for general values of $\beta$, but after applying some cases on $\beta$, solution can be obtained. It can be satisfied for two cases of $a$, which are used in Eq. (\ref{3.3}). For sufficiently large $a$, we have $\beta<\frac{2}{3C_{1}}$ and $V_{0}=0$ and for sufficiently small $a$, we have $\beta>\frac{2}{3C_{1}}$.
\begin{itemize}
\item Case (i): $\beta\neq\frac{2}{3C_{1}}$
\end{itemize}
The solution of inflaton field is found as
\begin{equation}\nonumber
\phi(N)=\pm M_{p}\sqrt{\frac{6\beta}{C_{3}}} N+ \phi_{0},
\end{equation}
where $C_{3}=\frac{2Q+2-3C_{1}\beta}{2-3 C_{1}\beta}$, leading to
\begin{equation}\label{3.10}
N=\pm \frac{\phi-\phi_{0}}{M_{p}} \sqrt{\frac{C_{3}}{6\beta}},
\end{equation}
now by putting the values of $N$ taking $V_{0}=0$ in Eq. (\ref{3.7}),
we get effective potential as
\begin{equation}\label{P}
V(\phi)=\frac{\dot{\phi_{0}}^{2}}{2\beta} \bigg(1+C_{2}Q-\beta \bigg) \exp\bigg(\mp \sqrt{6C_{3}\beta}\bigg(\frac{\phi-\phi_{0}}{M_{p}}\bigg)\bigg).
\end{equation}
\begin{figure} \centering
\epsfig{file=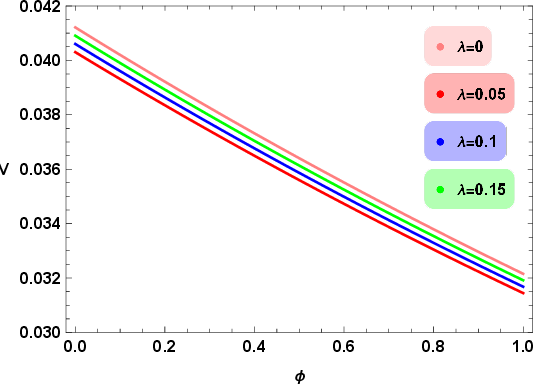, width=0.45\linewidth,
height=2.1in}\epsfig{file=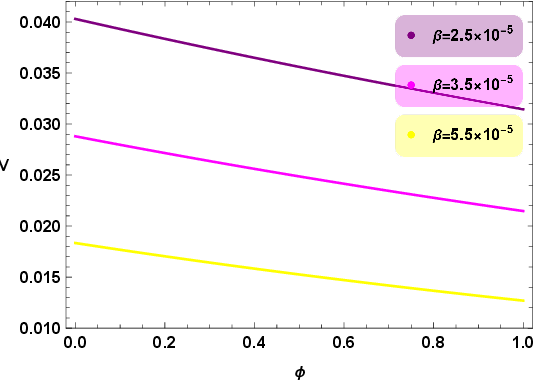, width=0.45\linewidth,
height=2.1in} \caption{$\phi-V$ trajectories are plotted
for $\kappa=1,~\gamma=\frac{4}{3},~Q= 2\times10^2,~M^{2}_{p}=1, ~\phi_{0}=0.01, ~\dot{\phi_{0}}=0.0001$ and $\lambda=0, 0.05, 0.1, 0.15$ (left panel) and $\beta=2.5\times10^{-5}, 3.5\times10^{-5}, 5.5\times10^{-5}$ (right panel).}\epsfig{file=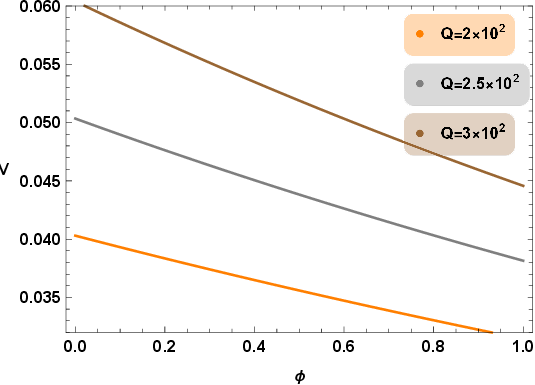, width=0.50\linewidth}
\caption{$\phi-V$ trajectories for $\kappa=1, \lambda=0.05, \gamma=\frac{4}{3}, \beta=2.5\times10^{-5}, M^{2}_{p}=1, \phi_{0}=0.01, \dot{\phi_{0}}=0.0001 $ and $Q= 2\times10^2, 2.5\times10^2, 3\times10^2$.}
\end{figure}
The above expression of potential is being plotted against inflaton field in Figs. \textbf{1} and \textbf{2} for different values of the involved model parameters. In our work, we mainly focus on checking the effects of Rastall parameter $\lambda$, dissipation factor $Q=Q_0$, and CR parameter $\beta$. The left plot of Fig. \textbf{1} shows that $V(\phi)$ has the same decaying trend versus $\phi$. It is directly proportional to $\lambda$ so
maximum value of potential increases with an increase in $\lambda$. The results are in good compatibility with GR curve plotted for $\lambda=0$. The right plot of Fig. \textbf{1} is plotted for different values of $\beta$. It is noted that $V(\phi)$ has an inverse relation with $\beta$, it decreases with an increase in the value of $\beta$. Figure \textbf{2} shows the effects of dissipation factor on the expression of $V(\phi)$. It can be seen that potential increases as the value of $Q$ increases.

Using the above expression of $N(\phi)$, One can easily obtain $\rho_r=TS$ in terms of $\phi$ from Eq. (\ref{3.3}). The expression of $\rho_r=TS$ is being plotted in Figs. \textbf{3} and \textbf{4} versus $\phi$ for different values of $\lambda$, $\beta$ and $Q$, respectively. The entropy density
has the same behavior as of $V(\phi)$. It increases as $\lambda$ increases (left plot of Fig. \textbf{3}) and decreases as $\beta$ increases (right plot of Fig. \textbf{3}). In Fig. \textbf{4}, we can see that firstly $S(\phi)$ increases with increasing values of $Q$ up to $2.5\times10^{2}$ but for larger values of $Q$, $S(\phi)$ drops down. The trajectories of $S-\phi$ are well-fitted with GR ($\lambda=0$). Further we conclude that the less value of potential is required to obtain inflation in RTG and less entropy density is produced as compared to GR.
\begin{figure} \centering
\epsfig{file=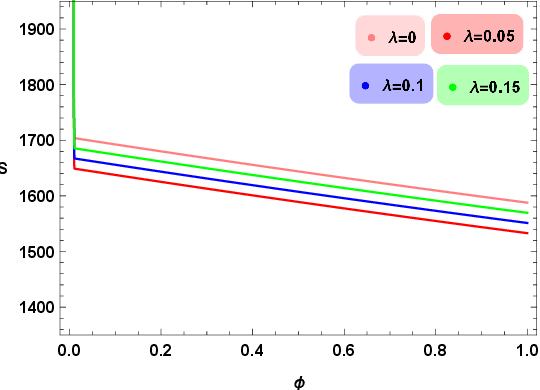, width=0.45\linewidth,
height=2.1in}\epsfig{file=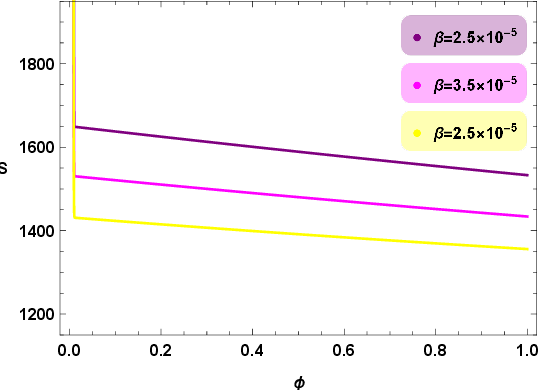, width=0.45\linewidth,
height=2.1in} \caption{$\phi-S$ trajectories are plotted
for $\kappa=1, \gamma=\frac{4}{3},\rho_{r_{0}} =0.001, Q= 2\times10^2, M^{2}_{p}=1, V_{0}=0.05, \phi_{0}=0.01, \dot{\phi_{0}}=0.0001, T=5.47\times10^{-5}$ and $\lambda=0, 0.05, 0.1, 0.15$ (left panel) and $\beta=2.5\times10^{-5}, 3.5\times10^{-5}, 5.5\times10^{-5}$ (right panel).}\epsfig{file=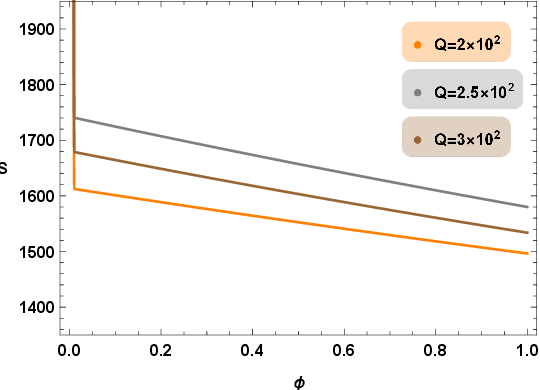, width=0.50\linewidth}
\caption{$\phi-S$ trajectories for $\kappa=1, \lambda=0.05, \gamma=\frac{4}{3},\rho_{r_{0}} =0.001, \beta=2.5\times10^{-5}, M^{2}_{p}=1, V_{0}=0.05, \phi_{0}=0.01, \dot{\phi_{0}}=0.0001, T=5.47\times10^{-5}$ and $Q= 2\times10^2, 2.5\times10^2, 3\times10^2$.}
\end{figure}

Now by solving Eq. (\ref{3.9}) taking $\rho_{r_0}=0$, this leads to the fact that the radiation are only present due to the dissipation,
otherwise there is no radiation. This case specially works for $\beta\ll1$, where the term $a^{-\frac{6\beta}{C_1}}$ (which appears in
Eq. (\ref{3.3})) is much lesser than $a^{-\frac{4}{C_{1}}}$ term. So, we can find solution of Eq. (\ref{3.9}) as
\begin{equation}\label{3.12}
\phi-\phi_{0}=\mp\sqrt{\frac{2 M_{p}^{2}}{3\beta C_{3}}} \sinh^{-1}\bigg(\sqrt{ \frac{\dot{\phi_{0}}^{2}C_{3}}{2\beta V_{0}}}e^{-\frac{3\beta}{C_1} N}\bigg),
\end{equation}
which shifted to
\begin{equation}\label{3.13}
N(\phi)=-\frac{C_1}{6\beta}\log\bigg(\frac{2\beta V_{0}}{\dot{\phi_{0}}^{2}C_{3}} \sinh^{2}\bigg(\mp\sqrt{\frac{3\beta C_{3}}{2 M_{p}^{2}}}(\phi-\phi_{0})\bigg)\bigg).
\end{equation}
So from Eq. (\ref{3.7}) and using above equation, potential can be obtained as
\begin{equation}\label{3.14}
V(\phi)=V_{0}\bigg(1+\frac{1+C_{2}Q-\beta}{C_{3}}\sinh^{2}\bigg(\mp\sqrt{\frac{3\beta C_{3}}{2 M_{p}^{2}}}(\phi-\phi_{0})\bigg)\bigg).
\end{equation}
This work is only valid for $\beta\neq\frac{2}{3C_{1}}$, even though radiation energy density has a pole at the value $3C_{1}\beta-2$.

\begin{itemize}
\item Case (ii): $\beta=\frac{2}{3C_{1}}$
\end{itemize}
One can find radiation energy density as follows
\begin{equation}\label{3.15}
\rho_{r}=\bigg(\rho_{r_{0}}+\frac{3Q \dot{\phi_{0}}^{2}}{C_{1}}\bigg)a^{-\frac{4}{C_{1}}}=\bigg(\rho_{r_{0}}+\frac{3Q \dot{\phi_{0}}^{2}}{C_{1}}\bigg)e^{-\frac{4N}{C_{1}}},
\end{equation}
which leads to
\begin{equation}\label{3.16}
3M_{p}^{2}H^{2}=\rho=C_{1}\bigg(\frac{3}{4}\dot{\phi_{0}}^{2}\bigg(1+Q\bigg(\frac{4N}{C_{1}}+1\bigg)\bigg)+\rho_{r_{0}}\bigg)
e^{-\frac{4N}{C_{1}}}+V_{0}.
\end{equation}
\begin{figure}\centering
\epsfig{file=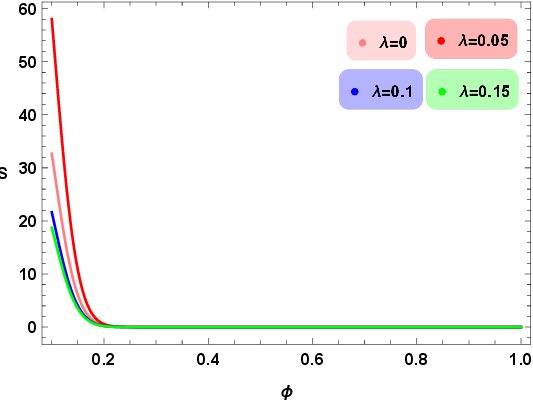, width=0.45\linewidth,
height=2.1in}\epsfig{file=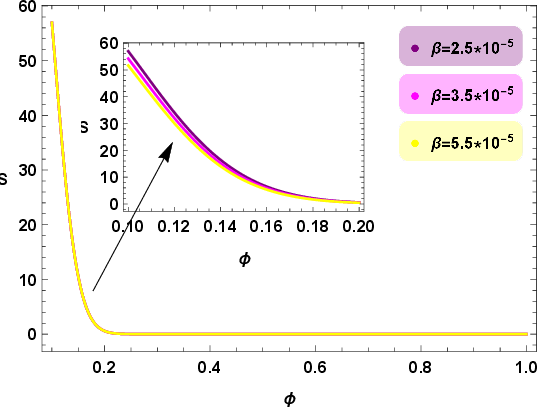, width=0.45\linewidth,
height=2.1in} \caption{$S-\phi$ trajectories are plotted
for $\kappa=1, \gamma=\frac{4}{3},\rho_{r_{0}} =5, Q= 2\times10^2, M^{2}_{p}=1, V_{0}=1\times10^{-4}, \phi_{0}=1\times10^{-4}, \dot{\phi_{0}}=0.5, T=5.47\times10^{-1}$ and $\lambda=0, 0.05, 0.1, 0.15$ (left panel) and $\beta=2.5\times10^{-5}, 3.5\times10^{-5}, 5.5\times10^{-5}$ (right panel).}\epsfig{file=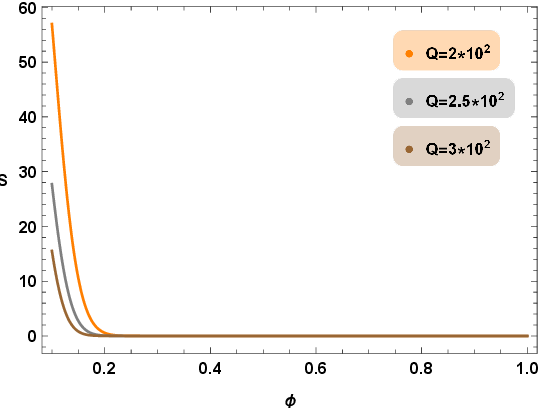, width=0.50\linewidth}
\caption{$S-\phi$ trajectories for $\kappa=1, \lambda=0.05, \gamma=\frac{4}{3},\rho_{r_{0}} =5, \beta=2.5\times10^{-5}, M^{2}_{p}=1, V_{0}=1\times10^{-4}, \phi_{0}=1\times10^{-4}, \dot{\phi_{0}}=0.5, T=5.47\times10^{-1}$ and $Q= 2\times10^2, 2.5\times10^2, 3\times10^2$.}
\end{figure}
The entropy density in this case follows the same trend as in the previous case. In the left plot of Fig. \textbf{5}, we can see that entropy density is higher for $\lambda=0.05$, which is not the case in Fig. \textbf{3} (left plot). The entropy density is less than GR for all other non-zero values of $\lambda$. The right plot shows that $S(\phi)$ is not much sensitive for values of $\beta$. The trajectories for all three values of $\beta$ are approximately overlapped and in the zoom graph it is seen to be decaying with increasing values of $\beta$. Figure \textbf{6} justifies the decreasing behavior of entropy density with increasing dissipation constant.
\begin{figure}\centering
\epsfig{file=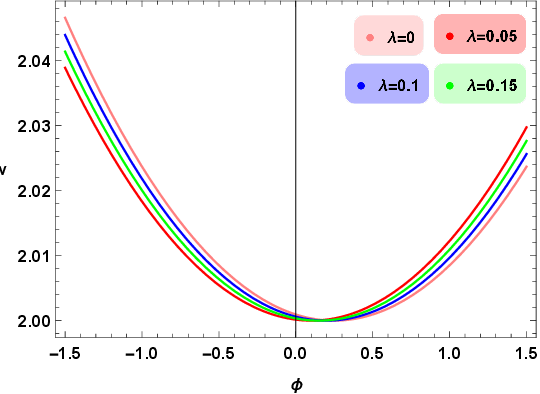, width=0.45\linewidth,
height=2.1in}\epsfig{file=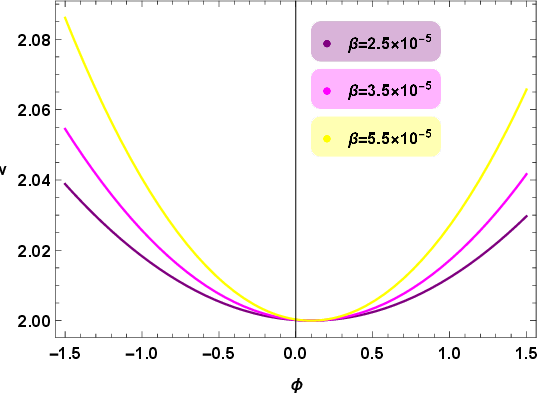, width=0.45\linewidth,
height=2.1in} \caption{$ V-\phi$ trajectories are plotted
for $\kappa=1, \gamma=\frac{4}{3}, Q= 2\times10^2, M^{2}_{p}=1, \phi_{0}=0.1, V_{0}=2$ and $\lambda=0, 0.05, 0.1, 0.15$ (left panel) and $\beta=2.5\times10^{-5}, 3.5\times10^{-5}, 5.5\times10^{-5}$ (right panel).}\epsfig{file=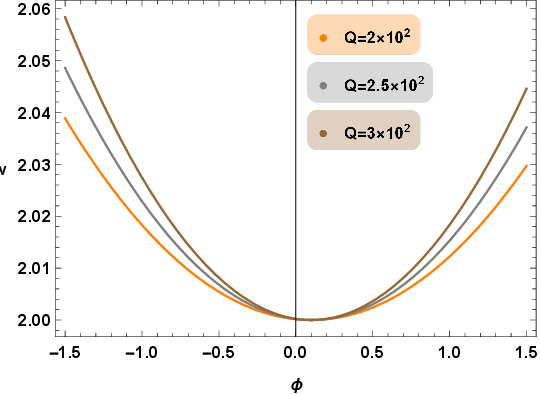, width=0.50\linewidth}
\caption{$ V-\phi$ trajectories for $\kappa=1, \lambda=0.05, \gamma=\frac{4}{3}, \beta=2.5\times10^{-5}, M^{2}_{p}=1, \phi_{0}=0.01, V_{0}=2 $ and $Q= 2\times10^2, 2.5\times10^2, 3\times10^2$.}
\end{figure}
Figures \textbf{7} and \textbf{8} are depicting the nature of obtained potential versus inflaton field. The results in this case are interesting as compared to previous case. The $V(\phi)$ has an oscillating behavior, goes down to its extreme minimum value and then push up to get an extreme value in each graph. The left plot shows an opposite trend to Fig. \textbf{1} (left plot), we require more amount of potential in RTG to produce inflation as compared to GR. Similarly the other graph (right plot of Fig. \textbf{7}) shows an opposite behavior as compared to previous case (Fig. \textbf{1}, left plot). The potential increases for increasing values of $\beta$ and $Q$.

Since $Q>0$, the energy density is always positive so for $\beta=\frac{2}{3C_1}$. We can calculate $\phi=\phi(N)$ as
\begin{equation}\label{3.17}
\phi(N)=\pm\sqrt{3}\dot{\phi_{0}}M_{p}\int\frac{e^{-\frac{2N}{C_{1}}}}{\sqrt{C_{1}\bigg(\frac{3}{4}\dot{\phi_{0}}^{2}\bigg(1+Q
\bigg(\frac{4N}{C_{1}}+1\bigg)\bigg)+\rho_{r_{0}}\bigg)e^{-\frac{4N}{C_{1}}}+V_{0}}}.
\end{equation}
For $V_{0}=0$, it turn out to be
\begin{equation}\label{3.18}
\phi(N)=\pm\frac{2M_{p}}{\sqrt{3}Q\dot{\phi_{0}}}\sqrt{C_{1}\bigg(\frac{3}{4}\dot{\phi_{0}}^{2}\bigg(1+Q\bigg(\frac{4N}{C_{1}}+1\bigg)\bigg)
+\rho_{r_{0}}\bigg)}+\phi_{0}.
\end{equation}
From Eqs. (\ref{3.7}) and (\ref{3.18}), we have
\begin{equation}\nonumber
V\propto\bigg(\exp\bigg(-Q \frac{(\phi-\phi_{0})^{2}}{M_{p}}\bigg)\bigg)^{\frac{3\beta}{2C_1}}.
\end{equation}
Note that the term $a^{-6\beta}$ appearing in Eq. (\ref{3.3}) becomes negative for
$Q>0$ and $\beta>\frac{2}{3C1}$, while the total energy density
remains positive.

\section{Inflationary Observables}

\subsection{Case(i): For $V_{0}=0$}

For small limit on $\beta$, it is the easiest regime where evolution
of inhomogeneities can be investigated. Therefore deviation from SR evolution
is not significant enough. In this regard, the energy density of our cosmos should
be dominated over the energy density of scalar field $\phi$. Similarly, the evolution of
radiation should also dominate over the dissipation term (i.e., the $a^{-\frac{6\beta}{C_1}}$ term).
The assumption $V_{0}=0$ shifts the current model to power-law model with WI.

The scalar power spectrum under CR limit is given as \cite{50}
\begin{equation}\label{4.1}
P_{\xi}=\bigg(\bigg(\frac{Q(1+\frac{3}{2}\beta)+1}{1+Q}\bigg)\frac{H}{\dot{\phi}}\bigg)^{2}
\frac{\Gamma_{R}(c+\frac{3}{2})}{\Gamma_{R}(\frac{3}{2})}
(2\Gamma T)\bigg(\frac{2\nu}{z^{2}}\bigg)^{3c}\sqrt{\frac{\Pi}{32\nu}},
\end{equation}
where $z=\frac{\kappa}{aH}$ and $z\rightarrow z\sqrt{1-3\beta}$, also we consider that $Q=\Gamma=$constant (does not depend
on temperature), $c=\frac{2(1+Q)\beta}{1-3\beta}\simeq(2\beta(1+Q))$, $\nu=\frac{3}{2}(1+Q)$
and $\Gamma_{R}$ is the gamma function. Using the expression of $H$ and $\dot{\phi}$, the power spectrum is turn out to be
\begin{eqnarray}\nonumber
P_{\xi}&=&\bigg(\frac{Q(1+\frac{3}{2}\beta)+1}{1+Q}\bigg)^{2}
\bigg(\frac{\phi _0^2}{6 \beta} \bigg(\frac{2Q+2-3C_{1}\beta}{2-3C_{1}\beta}\bigg)e^{-\frac{6\beta}{C_1} N}+\frac{\rho_{r_{0}}C_{1}}{3}e^{-\frac{4N}{C_{1}}}\bigg)
\\\label{4.2} & \times & (2\Gamma T)\frac{\Gamma_{R}(c+\frac{3}{2})}{\Gamma_{R}(\frac{3}{2})}
\left(\frac{2\nu}{z^{2}}\right)^{3c}\sqrt{\frac{\Pi}{32\nu}}.
\end{eqnarray}
\begin{figure} \centering
\epsfig{file=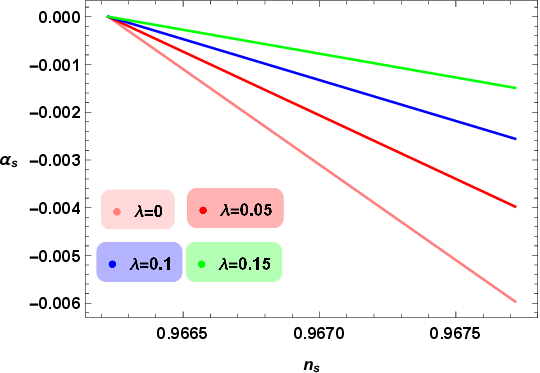, width=0.45\linewidth,
height=2.1in}\epsfig{file=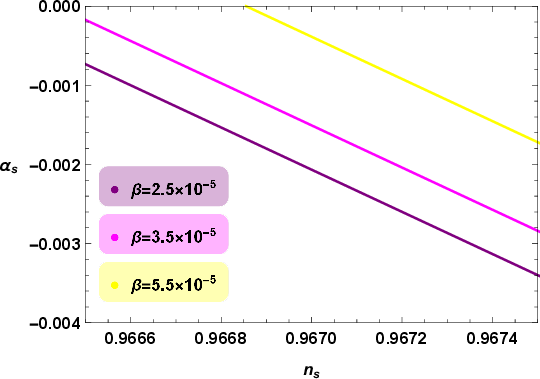, width=0.45\linewidth,
height=2.1in} \caption{$n_{s}-\alpha_{s}$ trajectories are plotted
for $\kappa=1, \gamma=\frac{4}{3},\rho_{r_{0}} =0.001, Q= 2\times10^2, \phi_{0}=0.001$ and $\lambda=0, 0.05, 0.1, 0.15$ (left panel) and $\beta=2.5\times10^{-5}, 3.5\times10^{-5}, 5.5\times10^{-5}$ (right panel).}\epsfig{file=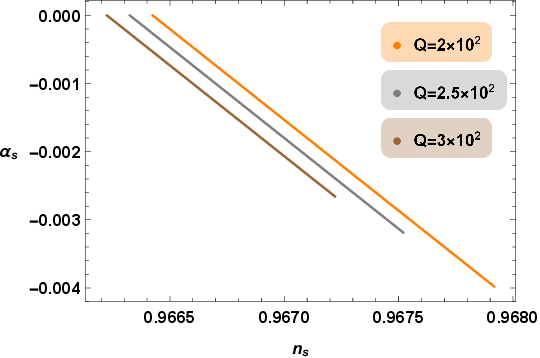, width=0.50\linewidth}
\caption{$n_{s}-\alpha_{s}$ trajectories for $\kappa=1, \lambda=0.05, \gamma=\frac{4}{3},\rho_{r_{0}} =0.001, \beta=2.5\times10^{-5}, \phi_{0}=0.001$ and $Q= 2\times10^2, 2.5\times10^2, 3\times10^2$.}
\end{figure}
\begin{figure} \centering
\epsfig{file=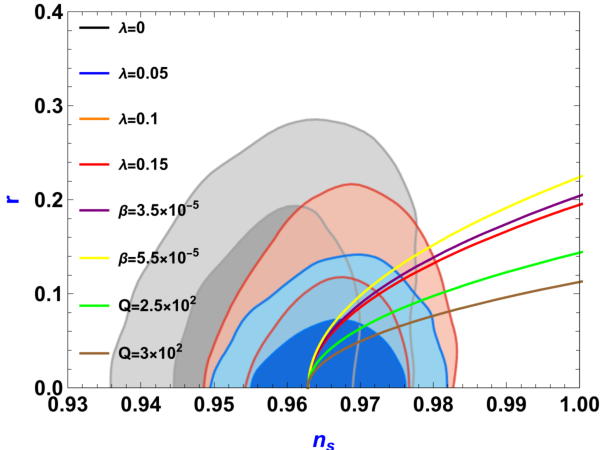, width=0.45\linewidth,
height=2.1in}\epsfig{file=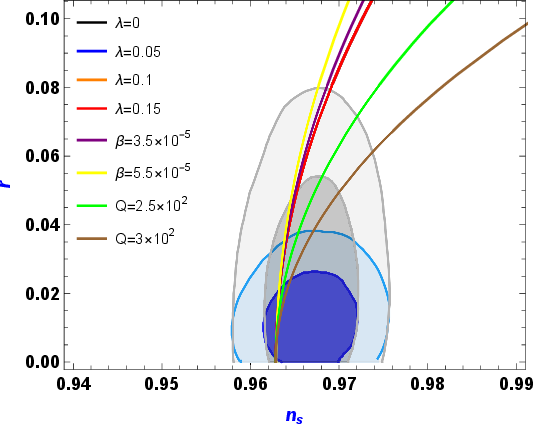, width=0.45\linewidth,
height=2.1in} \caption{$n_{s}-r$ trajectories are plotted
for $\kappa=1, \gamma=\frac{4}{3},\rho_{r_{0}} =0.001, Q= 2\times10^2, \phi_{0}=0.0001, T=5.47\times10^{-5}, \nu=0.008$ and $\lambda=0, 0.05, 0.1, 0.15$ (left panel) and $\beta=2.5\times10^{-5}, 3.5\times10^{-5}, 5.5\times10^{-5}$ (right panel).}
\end{figure}
The scalar spectral index and its running can be evaluated as
\begin{eqnarray}\label{4.3}
n_{s}-1&=&\frac{d\ln P_{\xi}}{d\ln k}=1-3 \beta -\frac{6 \beta  \left(2-3 \beta  C_1\right){}^2 e^{6 \beta  N} \rho _{r_0}}{\phi _0^2 e^{\frac{4 N}{C_1}} \left(3 \beta  C_1-2 (Q+1)\right)+2 \beta  C_1 \left(3 \beta  C_1-2\right) e^{\frac{6 \beta}{C_1}  N} \rho _{r_0}},\\
\label{4.5}
\alpha_{s}&=&\frac{d\ln n_{s}}{d\ln k}=-\frac{12 \beta  \phi _0^2 \left(3 \beta  C_1-2\right)^3 \rho _{r_0} e^{\frac{4 N}{C_1}+\frac{6 \beta}{C_1}  N} \left(3 \beta  C_1-2 (Q+1)\right)}{C_1 \left(\phi _0^2 e^{\frac{4 N}{C_1}} \left(-3 \beta  C_1+2 Q+2\right)-2 \beta  C_1 \left(3 \beta  C_1-2\right) e^{\frac{6 \beta}{C_1}  N} \rho _{r_0}\right)^2}.
\end{eqnarray}
During WI, thermal bath does not affect the tensor perturbations of space-time metric. Thus the standard definition of tensor power spectrum at horizon crossing is $P_{T}=\frac{8}{M_{p}}(\frac{H}{2\pi})^{2}$. These scalar and tensor power spectra lead us to generate another significant parameter of inflation dubbed as tensor-to-scalar ratio, which can be constrained directly by observational data. It can be calculated as
\begin{equation}\label{4.4}
r=\frac{P_{T}}{P_{\xi}}=\left(\frac{Q+1}{\left(1+\frac{3 \beta}{2}\right) Q+1}\right)^2 \frac{\Gamma_R \left(\frac{3}{2}\right)}{\Gamma_R \left(c+\frac{3}{2}\right)}\bigg(\frac{\sqrt{32 \nu }}{\sqrt{\pi}(2\nu)^{3 c}}\bigg)\frac{\phi _0^2 e^{-\frac{6 \beta}{C_1}N}}{3(Q+1) T \sqrt{H}},
\end{equation}
where $H=\frac{\phi^2_0e^{-\frac{6\beta}{C_1}N}\left(-3\beta C_1+2Q+2\right)}{(6\beta)\left(2-3\beta C_1\right)}+\frac{1}{3}C_1 e^{-\frac{4N}{C_1}}\rho _{r_0}$.

In left and right plots of Fig. \textbf{9}, $\alpha_{s}$ is plotted versus $n_{s}$ by varying $\lambda$ and $\beta$, respectively. The left plot shows that an allowed range for $\alpha_{s}$ is always obtained for all positive values of $\lambda$. The right plot constraint the interval of $\beta$, it should lie in the range $2.5\times10^{-5}\leq\beta\leq4\times10^-5$ to attain best fit results with Planck data. The yellow curve for $\beta=5.5\times10^{-5}$ in the right plot justifies the incompatibility of $\alpha_{s}$ with Planck data and the same holds for higher values of $\beta$. In Fig. \textbf{10}, $\alpha_{s}-n_s$ trajectories for different values of $Q$ are in good compatibility with data and it is checked that this compatibility holds for $Q\leq10^{13}$ (which is too large it is preferable to take less values of dissipation constant). Further it can be seen that with increasing dissipation, the scale of $n_s$ reduces from the allowed scale.

In the left plots of Figs. \textbf{11} and \textbf{12}, gray contours show the observational data of Planck 2013, red contours show Planck TT+lowP (2015) and blue contours show Planck TT, TE, EE+lowP (2015). While the right graphs are plotted to constraint the $r-n_s$ trajectories with Planck 2018 data (gray contours) and BICEP/Keck 2021 (blue contours) \cite{bicep}. It can be seen from both plots of Fig. \textbf{11} that the assumed CR inflationary model for $V_0=0$ is in good compatibility with Planck data of the year 2013, 2015, 2018 and BICEP/Keck 2021 for different values of $\lambda,~Q,~\beta$.

\subsection{Case(ii): For $V_{0}\neq0$}

The potential $V(\phi)$ is more general as $V_{0}$ is considered as essential part of energy density during CRWI, then the above work for perturbation can be re-evaluated and therefore more generalized value of $\beta$ can be found. In this case, $\epsilon\propto a^{-6\beta}$,
as SR parameter $\epsilon$ is very small, that is why $\epsilon$ parameter can be fully neglected. The observable $n_{s}$, for non-zero
potential can be calculated as follows
\begin{equation}\label{4.6}
n_{s}-1=1-12\beta-\frac{3 \left(\frac{\phi _0^2 e^{-\frac{6 \beta}{C_1}  N} \left(-3 \beta  C_1+2 Q+2\right)}{3 \beta  C_1-2}-\frac{4}{3} e^{-\frac{4 N}{C_1}} \rho _{r_0}\right)}{2 \left(\frac{\phi _0^2 e^{-\frac{6 \beta}{C_1}  N} \left(-3 \beta  C_1+2 Q+2\right)}{6 \beta  \left(2-3 \beta  C_1\right)}+\frac{1}{3} C_1 e^{-\frac{4 N}{C_1}} \rho _{r_0}+V_0\right)}.
\end{equation}
Tensor-to-scalar ratio has the following form
\begin{equation}\label{4.7}
r=\left(\frac{Q+1}{\left(1+\frac{3 \beta }{2}\right) Q+1}\right)^2 \frac{\Gamma \left(\frac{3}{2}\right)}{\Gamma \left(c+\frac{3}{2}\right)}
\frac{\sqrt{32 \nu }}{\sqrt{\pi } (2 \nu )^{3 c}}\bigg(\frac{\phi _0^2 e^{-\frac{6 \beta}{C_1}  N}}{3 (Q+1) T \sqrt{H}}\bigg),
\end{equation}
where $H=\frac{\phi_0^2 e^{-\frac{6 \beta}{C_1} N} \left(-3\beta C_1+2 Q+2\right)}{6\beta \left(2-3 \beta  C_1\right)}+\frac{1}{3} C_1 e^{-\frac{4 N}{C_1}} \rho _{r_0}+V_0$.
\begin{figure} \centering
\epsfig{file=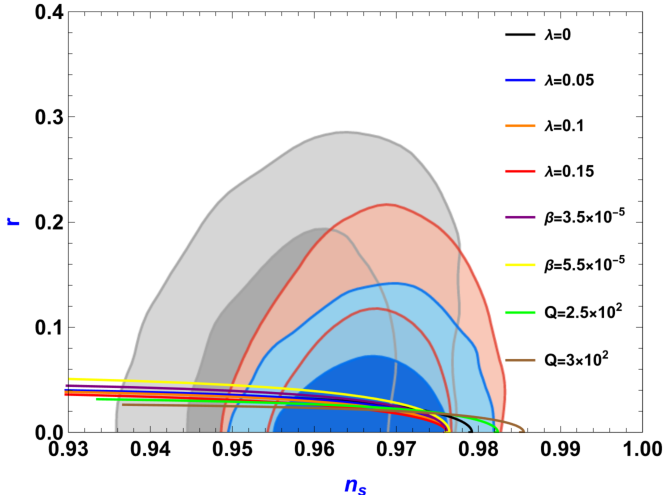, width=0.45\linewidth,
height=2.1in}\epsfig{file=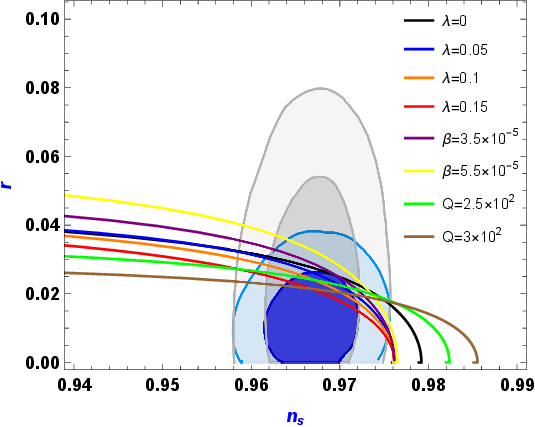, width=0.45\linewidth,
height=2.1in} \caption{$n_{s}-r$ trajectories are plotted
for $\kappa=1, \gamma=\frac{4}{3},\rho_{r_{0}} =0.001, Q= 2\times10^2, \phi_{0}=0.0001, T=5.47\times10^{-5}, \nu=0.01, V_{0}=2$ and $\lambda=0, 0.05, 0.1, 0.15$ (left panel) and $\beta=2.5\times10^{-5}, 3.5\times10^{-5}, 5.5\times10^{-5}$ (right panel).}
\end{figure}
From left and right plots of Fig. \textbf{12}, we can conclude that the WI model under CR condition for non-vanishing $V_0$ is fitted-well with Planck data (up to $2\sigma$ level)
and BICEP/Keck 2021 for different values of involved parameters. It is important to mention here that results are consistent with GR, shown by the black curve for $\lambda=0$.

The running of spectral-index is found to be
\begin{eqnarray}\nonumber
\alpha_{s}&=&12 \beta C^{-1}_1 \left(3 \beta  C_1-2\right) e^{\frac{4 N}{C_1}+\frac{6 \beta}{C_1}N} \bigg(\phi_0^2 \left(3 \beta  C_1-2 (Q+1)\right) \left(27 \beta^2 C_1 V_0 e^{\frac{4 N}{C_1}} + \left(2-3 \beta C_1\right)^2 \rho _{r_0}\right)\\\nonumber&+&24\beta V_0 \left(3\beta  C_1-2\right) e^{\frac{6 \beta}{C_1}  N} \rho _{r_0}\bigg)\bigg(\bigg(e^{\frac{4 N}{C_1}} \left(6\beta  V_0 \left(3 \beta C_1-2\right) e^{\frac{6 \beta}{C_1}  N}+\phi_0^2 \left(3\beta C_1-2 (Q+1)\right)\right)\\\label{4.8} &+&2\beta C_1 \left(3\beta  C_1-2\right) e^{\frac{6 \beta}{C_1} N} \rho _{r_0}\bigg)\bigg)^{-2}.
\end{eqnarray}
\begin{figure} \centering
\epsfig{file=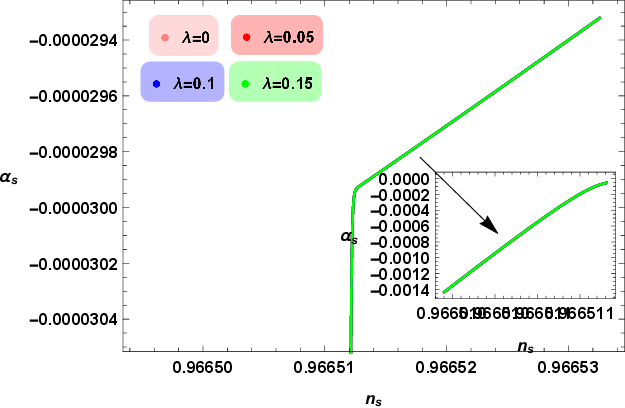, width=0.45\linewidth,
height=2.1in}\epsfig{file=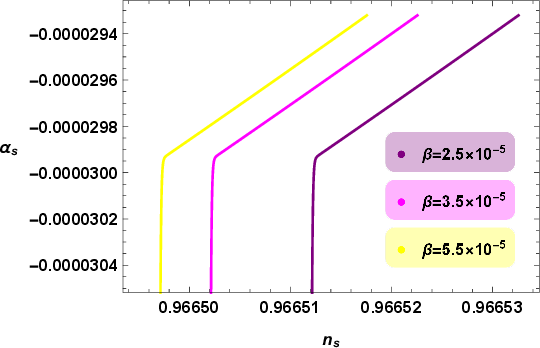, width=0.45\linewidth,
height=2.1in} \caption{$n_{s}-\alpha_{s}$ trajectories are plotted
for $\kappa=1, \gamma=\frac{4}{3},\rho_{r_{0}} =0.001, Q= 2\times10^2, \phi_{0}=0.0009, V_{0}=-3$ and $\lambda=0, 0.05, 0.1, 0.15$ (left panel) and $\beta=2.5\times10^{-5}, 3.5\times10^{-5}, 5.5\times10^{-5}$ (right panel).}\epsfig{file=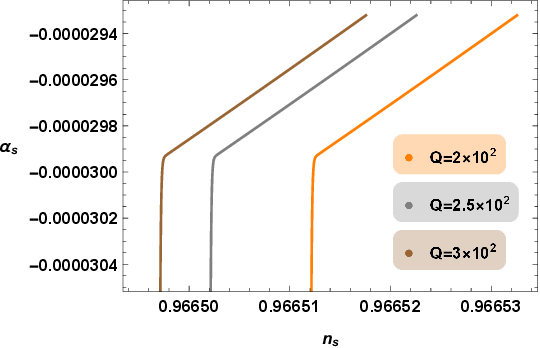, width=0.50\linewidth}
\caption{$n_{s}-\alpha_{s}$ trajectories for $\kappa=1, \lambda=0.05, \gamma=\frac{4}{3},\rho_{r_{0}} =0.001, \beta=2.5\times10^{-5}, \phi_{0}=0.0009, V_{0}=-3$ and $Q= 2\times10^2, 2.5\times10^2, 3\times10^2$.}
\end{figure}
In Figs. \textbf{13} and \textbf{14}, $n_s-\alpha_s$ trajectories are plotted by varying $\lambda, \beta$ and $Q$, respectively. The results are compatible with Planck data but $\alpha_s$ attain lowest values with respect to $n_s$.

\section{Concluding Remarks}

In this paper, we have constructed a simple and natural generalization of the CR inflationary models in GR to the case of Rastall gravity. We have evaluated the exact inflationary solutions under CR conditions within assumed gravity. In the entire work, the CR equation of motion has been assumed to be satisfied by the inflaton field.  Motivated by the paper \cite{36} since it gives a new approach to the construction of inflationary potential that is beneficial for the evolution of the field. Therefore, rather than using the field variable, we have constructed the results in terms of the number of e-folds as it appears to be a useful tool in WI. Hence, analytical expressions of scalar field $\phi$, its potential $V(\phi)$, and Hubble parameter $H$ are evaluated in terms of $N$. In the end, this approach proved to be consistent with the obtained analytical solutions in constant-roll inflation in GR.

Next, this study is expanded to the WI case, where the inflaton dissipates into relativistic degrees of freedom through dissipation coefficient $\Gamma$. During high dissipation regime where $\Gamma\gg3H$, we choose the simplest case $Q=\frac{\Gamma}{3H}$ (assuming $Q=$constant). Under this framework, we have found the expressions for $\phi,~V(\phi),~H(\phi)$ and the radiation energy density $\rho_{r}$. As it can be seen that the terms $a^{-\frac{6\beta}{C_1}}$ and $a^{-4}$ are included in the expression of $\rho_{r}$ as these terms come from redshifts like the kinetic term of inflaton field and pre-inflationary radiation, respectively. For $\beta>\frac{2}{3C_{1}}$ the term $a^{-\frac{6\beta}{C_1}}$ is negative. Even though the radiation energy density still decreases slower than the model with CI.

Furthermore, the origin of primordial inhomogeneities has been examined in two categories. The power spectrum of curvature perturbations with a slight deviation to scale invariance has been established for $\beta\ll1$, which implies little deviation from SR. For the case $\beta=\frac{2}{3C_{1}}$, we have $\frac{\rho_{r}C_{1}}{\dot{\phi}^{2}}\sim N$. From this result, the right-hand-side of the second Friedmann equation dominates significantly by radiation. Furthermore, we have evaluated the spectral-index $n_{s}$, tensor-to-scalar ratio $r$ and the running of spectral-index $\alpha_{s}$ taking $V_{0}=0$ and $V_{0}\neq0$, alternatively. The results for these cases are as follows:

\textbf{For $V_0=0$:}
\begin{itemize}
  \item The obtained potential is plotted against $\phi$ in Figs. \textbf{1} and \textbf{2} for different values of the involved model parameters. The left plot of Fig. \textbf{1} shows that $V(\phi)$ has decaying trend versus $\phi$. It is directly proportional to $\lambda$, and the results are in good compatibility with GR curve, which is plotted for $\lambda=0$. It is noted from the right plot of Fig. \textbf{1} that $V(\phi)$ has an inverse relation with $\beta$. Figure \textbf{2} shows that potential increases as the value of $Q$ increases.
  \item  The entropy density has the same behavior as of $V(\phi)$. It increases as $\lambda$ increases (left plot of Fig. \textbf{3}) and decreases as $\beta$ increases (right plot of Fig. \textbf{3}). In Fig. \textbf{4}, we can see that firstly $S(\phi)$ increases with increasing values of $Q$ up to $2.5\times10^{2}$ but for larger values of $Q$, $S(\phi)$ drops down. The trajectories of $S-\phi$ are well-fitted with GR ($\lambda=0$). In this case, we conclude that the less value of potential is required to obtain inflation in RTG and less entropy density is produced as compared to GR.
  \item The left plot of Fig. \textbf{9} shows that an allowed range for $\alpha_{s}$ is always obtained for all positive values of $\lambda$. The right plot constraint the interval of $\beta$, it should lie in the range $2.5\times10^{-5}\leq\beta\leq4\times10^-5$ to attain best fit results with Planck data. The yellow curve for $\beta=5.5\times10^{-5}$ in the right plot justifies the incompatibility of $\alpha_{s}$ with Planck data and the same holds for higher values of $\beta$. In Fig. \textbf{10}, $\alpha_{s}-n_s$ trajectories for different values of $Q$ are in good compatibility with data and it is checked that this compatibility holds for $Q\leq10^{13}$ (which is too large it is preferable to take less values of dissipation constant). Further it can be seen that with increasing dissipation, the scale of $n_s$ reduces from the allowed scale.
  \item It can be seen from both plots of Fig. \textbf{11} that the assumed CR inflationary model for $V_0=0$ is in good compatibility with Planck data published in 2013, 2015, 2018 and BICEP/Keck 2021 for different values of $\lambda, Q$ and $\beta$.
\end{itemize}

\textbf{For $V_0\neq0$:}
\begin{itemize}
  \item Figures \textbf{7} and \textbf{8} are depicting the nature of obtained potential versus inflaton field. The results in this case are interesting as compared to previous case. The $V(\phi)$ has an oscillating behavior, goes down to its extreme minimum value and then push up to get an extreme value in each graph. The left plot shows an opposite trend to Fig. \textbf{1} (left plot), we require more amount of potential in RTG to produce inflation as compared to GR. Similarly the other graph (right plot of Fig. \textbf{7}) shows an opposite behavior as compared to previous case (Fig. \textbf{1}, left plot). The potential increases for increasing values of $\beta$ and $Q$.
  \item  The entropy density in this case follows the same trend as in the previous case. In the left plot of Fig. \textbf{5}, we can see that entropy density is higher for $\lambda=0.05$, which is not the case in Fig. \textbf{3} (left plot). The entropy density is less than GR for all other non-zero values of $\lambda$. The right plot shows that $S(\phi)$ is not much sensitive for values of $\beta$. The trajectories for all three values of $\beta$ are approximately overlapped and in the zoom graph it is seen to be decaying with increasing values of $\beta$. Figure \textbf{6} justifies the decreasing behavior of entropy density with increasing dissipation constant.
  \item From left and right plots of Fig. \textbf{12}, we can conclude that the WI model under CR condition for non-vanishing $V_0$ is fitted-well with Planck data up to $2\sigma$ level and BICEP/Keck 2021 for different values of involved parameters. It is important to mention here that results are consistent with GR, shown by the black curve for $\lambda=0$.
  \item In Figs. \textbf{13} and \textbf{14}, $n_s-\alpha_s$ trajectories are plotted by varying $\lambda, \beta$ and $Q$, respectively. The results are compatible with Planck 2013, 2015, 2018 data and BICEP/Keck 2021 data but $\alpha_s$ attain lowest values with respect to $n_s$.
\end{itemize}

Finally, we conclude that the CR technique is an efficient technique to find the exact inflationary solutions to the gravitational field equations. This paper is the first approach to finding the exact solutions of the model with WI by utilizing this technique of CR within RTG. This work might produce interesting results for variable dissipative factor as discussed in the literature for the inflation inspired by SR conditions. Since it is the first study, we use the simplest case of constant $Q$. Using the framework of modified theories of gravities, this technique can produce fruitful results in inflation like in $f(R)$ and $f(T)$ theories \cite{35, awad}.

\vspace{0.25cm}
\textbf{Data Availability Statements}\\

All data generated or analyzed during this study are included in this published article.


\begin{thebibliography}{40}

\bibitem{1} A. A. Starobinsky, Phys. Lett. B \textbf{91}, 99 (1980).

\bibitem{2} A. H. Guth, Phys. Rev. D \textbf{23}, 347 (1981)

\bibitem{3}  K. I. Maeda, and K. Yamamoto, J. Cosmol. Astropart. Phys. \textbf{018}, 12 (2013).

\bibitem{4} A. A. Abolhasani, R. Emami, and H. Firouzjahi, J. Cosmol. Astropart. Phys. \textbf{016}, 05 (2014), .

\bibitem{5} S. Alexander et al., J. Cosmol. Astropart. Phys. \textbf{005}, 05 (2015).

\bibitem{6} K. Saaidi, A. Mohammadi, and T. Golanbari, Adv. High Energy Phys. 926807 (2015) .

\bibitem{7} N. Nazavari, A. Mohammadi, Z. Ossoulian, and K. Saaidi, Phys. Rev. D \textbf{93}, 123504 (2016).

\bibitem{8} K. Sayar, A. Mohammadi, L. Akhtari, and K. Saaidi, Phys. Rev. D \textbf{95}, 023501 (2017).

\bibitem{10} A. Mohammadi, K. Saaidi, and H. Sheikhahmadi, Phys. Rev. D \textbf{100}, 083520 (2019).

\bibitem{11} A. Mohammadi et al.,  Chin. Phys. C \textbf{44}, 095101 (2020).

\bibitem{12} T. Golanbari, A. Mohammadi, and K. Saaidi, Phys. Dark Univ. \textbf{27}, 100456 (2020).

\bibitem{13} Y. Akrami et al. (Planck), Planck 2018 results. Astron. Astrophys. \textbf{641}, A10 (2020).

\bibitem{14} A. Riotto, ICTP Lect. Notes Ser. \textbf{14}, 317 (2003).

\bibitem{15} A. D. Linde, Phys. Scripta \textbf{117}, 40 (2005).

\bibitem{16} S. Weinberg, \textit{Cosmology} (Oxford University Press, 2008).

\bibitem{17} D. Baumann, \textit{TASI Lecture on Inflation, in Physics of the large and the small, TASI 09, proceedings of the Theoretical Advanced Study} (2011).

\bibitem{18} A.D. Linde, Phys. Lett. B \textbf{129}, 117 (1983).

\bibitem{19} A.D. Linde, Phys. Rev. D \textbf{49}, 748 (1994).

\bibitem{20} A.D. Linde, Phys. Lett. B \textbf{108}, 389 (1982).

\bibitem{21} H. Motohashi, A.A Starobinsky and J. Yokoyama, J. Cosmo. Astropart. Phys. \textbf{1509}, 018 (2015).

\bibitem{47a} W.H. Kinney, Phys. Rev. D \textbf{72}, 023515 (2005).

\bibitem{47b} N.C. Tsamis and R.P. Woodard, Phys. Rev. D \textbf{69}(2004)084005.

\bibitem{22} J. Martin, H. Motohashi and T. Suyama, Phys. Rev. D \textbf{87}, 023514 (2013).

\bibitem{22a} M.H. Namjo, H. Firouzjati and M. Sasaki, Euro. Phys. Lett. \textbf{101}(2013)39001.

\bibitem{24} H. Motohashi and W. Hu, Phys. Rev. D \textbf{96}, 023502 (2017).

\bibitem{25} S. D. Odintsov and V. K. Oikonomou,  Phys. Rev. D \textbf{96}, 024029 (2017).

\bibitem{26} L. Anguelova, P. Suranyi and L. C. R. Wijewardhana,  J. Cosmol. Astropart. Phys. \textbf{02}, 004 (2018).

\bibitem{27} H. Motohashi and A. A. Starobinsky, Eur. Phys. J. C \textbf{77}, 538 (2017).

\bibitem{28} S. Nojiri, S. D. Odintsov and V. K. Oikonomou, Class. Quant. Grav. \textbf{34}, 245012 (2017).

\bibitem{29} A. Karam, L. Marzola, T. Pappas, A. Racioppi and K. Tamvakis,  J. Cosmol. Astropart. Phys. \textbf{05}, 011 (2018).

\bibitem{31} H. Motohashi, A. A. Starobinsky, Eur. Phys. J. C \textbf{77}, 538 (2017).

\bibitem{32} F. Cicciarella, J. Mabillard, M. Pieroni, J. Cosmol. Astropart. Phys. \textbf{01}, 024 (2018).

\bibitem{33} A. Ito, J. Soda, Eur. Phys. J. C \textbf{78}, 55 (2018).

\bibitem{34} A. Oliveros, H.E. Noriega, Internat. J. Modern Phys. D \textbf{28}, 1950159 (2019).

\bibitem{35} H. Motohashi, A. A. Starobinsky, J. Cosmol. Astropart. Phys. \textbf{11}, 025 (2019).

\bibitem{36} V. Kamali, M. Artymowski, M. R. Setare, J. Cosmol. Astropart. Phys. \textbf{07}, 002 (2020).

\bibitem{awad} A. Awad et al., J. Cosmol. Astropart. Phys. \textbf{07}, 026 (2018).

\bibitem{37}  M. Guerrero, D. Rubiera-Garcia, D. S. C. Gomez, Phys. Rev. D \textbf{102}, 123528 (2020).

\bibitem{38}  M. Shokri, J. Sadeghi, M. R. Setare, Ann. Physics \textbf{429}, 168487 (2021).

\bibitem{39} V. K. Oikonomou, arXiv:2106.10778.

\bibitem{40} M. Shokri, M.R. Setare, S. Capozziello, J. Sadeghi, Eur. Phys. J. Plus \textbf{137}, 1 (2022).

\bibitem{R1} C. Germani and T. Prokopec, Phys. Dark Uni. \textbf{18}, 6 (2017).

\bibitem{R2} Y. Gong, arXiv: 1707.09578.

\bibitem{R3} R. Saito, J. Yokoyama and R. Nagata, JCAP \textbf{0806}, 024 (2008).

\bibitem{41aa} A. Berera, Phys. Rev. Lett. \textbf{75}, 3218 (1995).

\bibitem{41b} D. H. Lyth and A. R. Liddle,
\textit{The primordial density perturbation: Cosmology, inflation and the origin of structure} (Cambridge University Press, 2009).

\bibitem{41} I. G. Moss and C. Xiong, J. Cosmol. Astropart. Phys. \textbf{04}, 007 (2007).

\bibitem{42} C. Graham and I. G. Moss, J. Cosmol. Astropart. Phys. \textbf{07}, 013 (2009).

\bibitem{43} R. O. Ramos and L. da Silva, J. Cosmol. Astropart. Phys. \textbf{03}, 032 (2013).

\bibitem{44} M. Bastero-Gil, A. Berera, and R. O. Ramos, J. Cosmol. Astropart. Phys. \textbf{07}, 030 (2011).

\bibitem{bulk} S. Weinberg, \textit{Gravitation and Cosmology} (Wiley, 1972).

\bibitem{45} J. P. Mimoso, A. Nunes and D. Povon, Phys. Rev. D \textbf{73}, 023502 (2006).

\bibitem{47} Q. Gao, Sci. China Phys. Mech. \textbf{61}, 070411 (2018).

\bibitem{R4} T.T. Gao, Eur. Phys. J. C \textbf{80}(2020)1013.

\bibitem{33n} F. S. N. Lobo, J. phys. Conference Series \textbf{600}, 012006 (2015).

\bibitem{34n} V. Faraoni, et al., \textit{The Landscape Beyond Einstein Gravity} (Springer, 2011)

\bibitem{35n} C. E. M. Batista, et al., Phys. Rev. D \textbf{85}, 084008 (2012).

\bibitem{36n} S. Nojiri and S.D. Odintsov, Phys. Lett. B \textbf{599}, 137 (2004).

\bibitem{37n} G. Allemand, et al., Phys. Rev. D \textbf{72}, 063505 (2005).

\bibitem{38n} O. Bertolami, et al., Phys. Rev. D \textbf{75}, 104016 (2007).

\bibitem{39n} P. Rastall, Phys. Rev. D \textbf{6}, 3357 (1972).

\bibitem{40n} M. Capone, V.F. Cardone and M.L. Ruggiaro, Phys. J. \textbf{222}, 012012 (2010).

\bibitem{41n} D. Das, S. Dutta and S. Chakraborty, Eur. Phys. J. C \textbf{78}, 810 (2018).

\bibitem{43n} R. Saleem and J. Hassan, Phys. Dark Uni. \textbf{28}, 100515 (2020).

\bibitem{me} R. Saleem and  I. Shahid, Phys. Dark Uni. \textbf{35}, 100920 (2022).

\bibitem{multi} M. Guerrero, D. Rubiera-Garcia, and D. S. C. Gamez, Phys. Rev. D \textbf{102}, 123528 (2020).

\bibitem{s2} M. R. Setare, A. Ravanpak, K. Bahari, and G. F. Fadakar, Int. J. Mod. Phys. D \textbf{30}, 2150116 (2021).

\bibitem{49} R. Jinno and K. Kaneta, Phys. Rev. D \textbf{96}, 043518 (2017).

\bibitem{50} O. Bertolami, et al., Phys. Rev. D \textbf{75}, 104016 (2007).

\bibitem{bicep} P. A. R. Ade, et al., Phys. Rev. Lett. \textbf{127}, 151301 (2021).

\end{thebibliography}
\end{document}